 \documentclass[manuscript]{acmart}
\AtBeginDocument{%
  }

\acmConference[CHI '25]{Workshop: AI Augmented Reasoning}{April 26-- May 01, 2025}{Yokohama, Japan}

\begin{document}

\title[DeBiasMe: De-biasing Human-AI Interactions with Metacognitive AIED (AI in Education) Interventions]{DeBiasMe: De-biasing Human-AI Interactions with Metacognitive AIED (AI in Education) Interventions}
\begingroup 
\renewcommand{\thefootnote}{} \footnote{This paper was presented at the 2025 ACM Workshop on Human-AI Interaction for Augmented Reasoning (AIREASONING-2025-01). This is the authors’ version for arXiv.}\endgroup

\author{Chaeyeon Lim}
\email{chaeyeon.lim.24@ucl.ac.uk}
\orcid{0009-0001-0058-6916}
\affiliation{
  \institution{UCL Interaction Centre}
  \city{London}
  \country{UK}
}

\begin{abstract}
While generative artificial intelligence (Gen AI) increasingly transforms academic environments, a critical gap exists in understanding and mitigating human biases in AI interactions, such as anchoring and confirmation bias. This position paper advocates for metacognitive AI literacy interventions to help university students critically engage with AI and address biases across the Human-AI interaction workflows. The paper presents the importance of considering (1) metacognitive support with deliberate friction focusing on human bias; (2) bi-directional Human-AI interaction intervention addressing both input formulation and output interpretation; and (3) adaptive scaffolding that responds to diverse user engagement patterns. These frameworks are illustrated through ongoing work on "DeBiasMe," AIED (AI in Education) interventions designed to enhance awareness of cognitive biases while empowering user agency in AI interactions. The paper invites multiple stakeholders to engage in discussions on design and evaluation methods for scaffolding mechanisms, bias visualization, and analysis frameworks. This position contributes to the emerging field of AI-augmented learning by emphasizing the critical role of metacognition in helping students navigate the complex interaction between human, statistical, and systemic biases in AI use while highlighting how cognitive adaptation to AI systems must be explicitly integrated into comprehensive AI literacy frameworks. 
\end{abstract}


\maketitle

\section{Introduction}

Generative artificial intelligence (Gen AI) tools, particularly text generators with LLMs, are rapidly transforming academic environments not just as productivity aids but as potential catalysts or barriers to critical thinking \cite{Hikmawati2025}.  The adoption of these tools for automation in various cognitive tasks (e.g., summarization, argument development, and research synthesis) rather than for augmentation poses cognitive risks in education by promoting over-reliance on AI  \cite{Zhai2024}, reinforcing existing beliefs \cite{ferrara2023}, and offloading higher-order thinking skills to AI systems \cite{lee2025, anthropic2025}.  Despite increasing AI literacy initiatives extending data and digital literacy frameworks to develop the skill set required for effective human-AI collaboration \cite{Long2020}, what is less discussed are cognitive biases or human errors in information processing and decision making in this new form of human-computer interactions \cite{AlonBarkat2022}. For example, automation bias can make students overly reliant on AI outputs \cite{Lee2004}, confirmation bias reinforces their existing beliefs \cite{Rosbach2024}, and anchoring bias locks them into initial AI-generated suggestions \cite{rastogi2022deciding}. Critically, these human biases interact with algorithmic and systemic biases embedded in AI systems and datasets \cite{Schwartz2022}. This interaction creates a harmful feedback loop that can amplify both types of biases: human biases influencing how users interact with AI systems, and AI biases potentially reinforcing human cognitive biases, negatively impacting the quality of learning and critical thinking \cite{Glickman2025} (See Fig. \ref{fig:bias_visualization}). 

\begin{figure}[h]
    \centering
    \includegraphics[width=\textwidth]{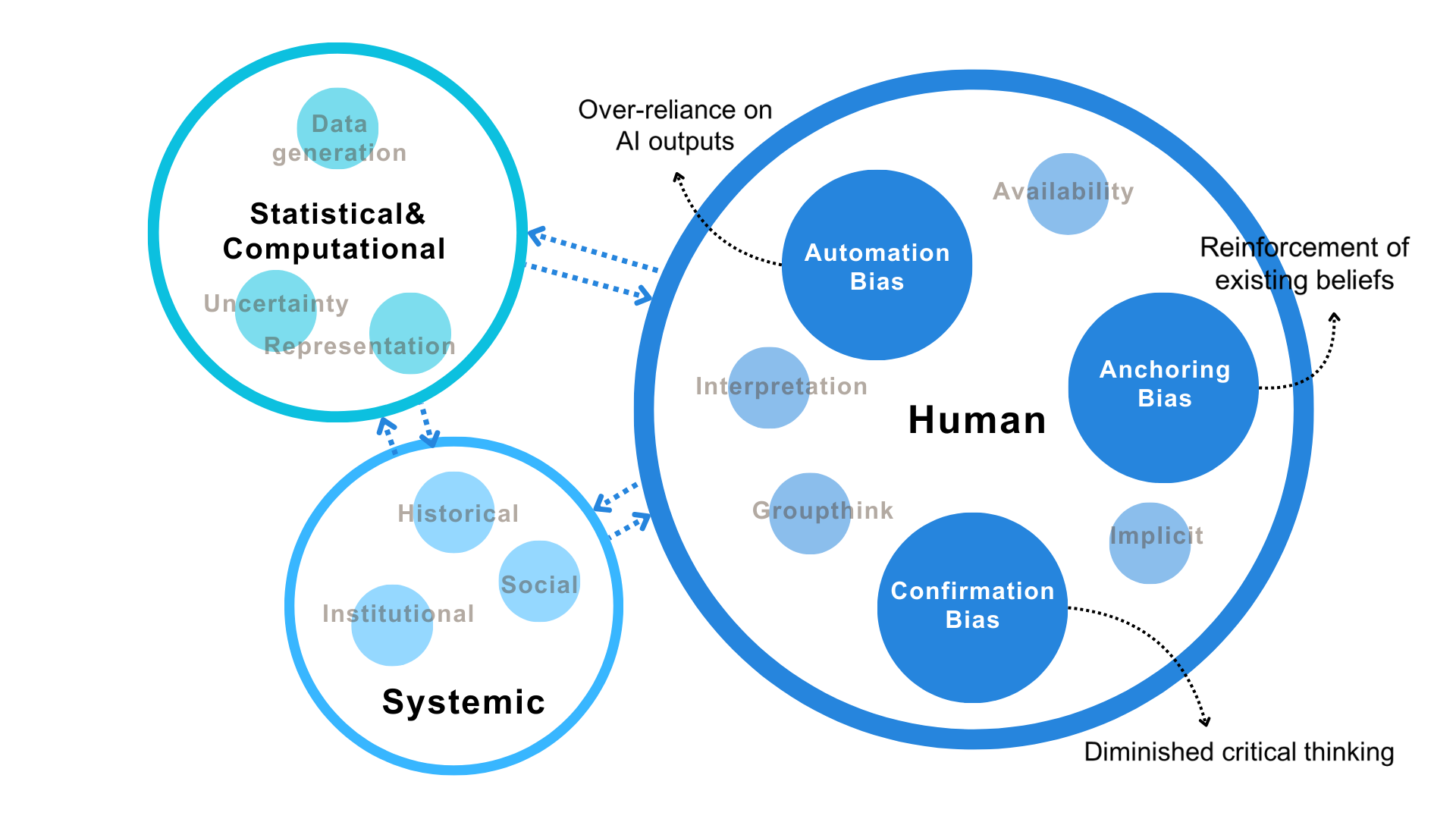} 
    \Description{}
    \caption{Interacting categories of bias in human-AI Interaction} 
    \label{fig:bias_visualization}
\end{figure}

To bring this discussion into focus, this position paper argues that AI literacy approaches should prioritize developing human bias awareness and metacognitive skills, the ability to monitor, evaluate, and regulate one's own learning and decision-making processes when interacting with AI tools \cite{sidra2024}. By fostering critical engagement and reflection, such tools can transform students from passive AI consumers to active collaborators with AI systems \cite{carnegie2025ai}. To illustrate these arguments, this paper presents work-in-progress project \textit{DeBiasMe}, a metacognitive AIED (AI in Education) literacy intervention for human and AI bias mitigation. This will be extended by discussing ongoing research and development directions and their educational and social implications. This position paper invites educators, researchers, designers, AI developers, and students to join in  \textit{DeBiasMe}'s exploration of the following questions (See Table \ref{tab:research_questions}) to advance AI literacy approaches that empower end-users in AI-augmented learning environments. 

\begin{table}[H]
  \caption{Research questions}
  \label{tab:research_questions}
  \begin{tabular}{p{4.5cm}p{10cm}}
    \toprule
    \textbf{Focused Area}  
& \textbf{Question}\\
    \midrule
    Deliberate Friction for Metacognitive Support& How can strategic integration of friction points in the human-AI workflow support critical thinking and metacognitive skills?\\
    Bias Interaction Patterns& How to address human bias at both the input (prompt formulation) and output stages (interpretation of AI responses) to help students navigate the complex interaction of human, systemic, and statistical biases?\\
    Individual factors& How to design scaffolding mechanism that adapts to individual factors involved in human-AI interaction?\\
    \bottomrule
  \end{tabular}
\end{table}

\section{Related Works}
\subsection{Technical gap: human bias in AI literacy tools}
Conventional Human-in-the-Loop interventions prioritize refining AI models with less attention to human involvement in both input and output stages of the interaction \cite{garcia2022}. As a result, existing AI literacy tools predominantly focus on algorithmic and systemic bias, with a lack of AI literacy interventions that address human biases (See Table \ref{tab:existing_solutions}). Further, existing tools are target domain-specific use cases or users with technical expertise, creating accessibility barriers that widen the digital divide and leave broader populations without effective means to develop critical AI competencies \cite{Ahmed2024}. 

\subsection{Theoretical gap: the role of metacognition in human-AI interaction}
Further, there is a gap in competency dimensions that current AI literacy is advocating for: the current emphasis on technical competencies in human intervention, such as prompt engineering, and ethical and responsible AI use, which is essential for human-AI collaboration beyond automation \cite{Chiu2024, Knoth2024}.  Bridging this gap in AI literacy frameworks requires AI literacy to move beyond first-level (operational understanding) to incorporate metacognitive skills by providing enhanced explainability and customizability in tool design and educational contents to support metacognitive demands in human-AI interaction \cite{sidra2024, tankelevitch2024}. However, their effectiveness depends on understanding the individual factors that shape how students engage with AI. While UNESCO competency frameworks for AI literacy capture various learning dimensions \cite{unesco2024}, mitigation of biases requires not only knowledge-based analytics but also attributional and affective dimensions involving individual factors, including the confidence \cite{chandra2022}, trust \cite{korber2019}, agency \cite{Berberian2012}, and anthropomorphism tendency \cite{chandra2022}, which needs to be further addressed to better support diverse learners \cite{ng2024}. 

\begin{table} 
  \caption{A survey of existing interventions}
  \label{tab:existing_solutions}
    \centering
    \begin{tabular}{p{2.5cm}p{3cm}p{2.5cm}p{2.5cm}p{3cm}}
    \toprule
    \textbf{Toolkit / Framework}& \textbf{Educational Focus \& Features}& \textbf{Bias Focus Area}& \textbf{Learning Interaction}& \textbf{Target Users}\\
    \midrule
    Google What-If Tool \cite{GoogleWhatIfTool}& Interactive visual interface to simulate changes in ML models& Fairness, AI bias& Jupyter Notebook / GUI& Undergraduate students, Data Science beginners\\
 AI Blindspot \cite{AIBlindspot}& Card-based prompts for uncovering cognitive and design biases& Cognitive bias& Workshop / Cards&High school and early college students\\
 D-BIAS Tool \cite{ghai2023} & Web-based visual tool showing dataset influence on AI decisions& Dataset-level bias& Web GUI&University students, HCI/UX learners\\
 Fairlearn \cite{Fairlearn2021}& Fairness metrics and disparity analysis in ML outcomes& Statistical bias& Coding / Dashboard&Technical undergrads, grad students\\
 IBM AI Fairness 360 \cite{AIFairness360} & Metrics \& bias mitigation strategies toolkit& Algorithmic bias& Jupyter notebooks&Grad students, ML-focused courses\\
 Perspective API \cite{PerspectiveAPI} & Detects toxic or biased language& Linguistic bias& API / Front-end integration&Communication/media studies\\
    OpenAI Moderation \cite{OpenAIModeration} & Filters harmful outputs from LLMs& Content  moderation& API / Embedded& AI system designers, CS learners\\
        \bottomrule
  \end{tabular}
\end{table}

\section{Conceptual Framework: towards learner-centered AIED}
To address identified technical and theoretical gaps, design solutions with a deliberate friction approach and a new theoretical framework for bi-directional intervention will be discussed as key foundations for the development of \textit{DeBiasMe}. 

\subsection{Deliberate friction in human-AI feedback loops}
While the design of AI as an automatic decision-making system tends to reduce cognitive load for users and minimize interruptions in user workflows  \cite{Nielsen1994}, design interventions known as "deliberate friction" can introduce intentional pauses and reflection opportunities \cite{Benedetti2023}. This reflective and value-sensitive approach facilitates active control and independent thinking over automation, helping users adjust their agency and trust levels \cite{bucinca2021} through strategic interventions  \cite{Sengers2005, Friedman2019}. These friction points serve as metacognitive triggers and knowledge scaffolding, reducing susceptibility to and amplification between cognitive and AI biases. 

\subsection{Bi-Directional human-AI collaboration}
Such design intervention can be effectively intergrated when users are perceived as active participants in AI interaction and decision-making processes \cite{shneiderman2020}. Adopting this framework in AI literacy allows reimagining the Human-in-the-Loop as a bi-directional process where humans can intervene in both input and output stages for close human-AI collaboration \cite{cukurova2024}. From this view, algorithmic and systemic biases are not imposed on users in a top-down manner, but end-users can address them with bias recognition and mitigation efforts. By empowering end-users with enhanced interpretability and transparency \cite{Gasevic2023}, this approach also contributes to both human-centered adoption of AI to human as well as human adaptation to AI systems  \cite{shen2024}. 

\begin{figure} 
    \centering
  \includegraphics[width=\textwidth]{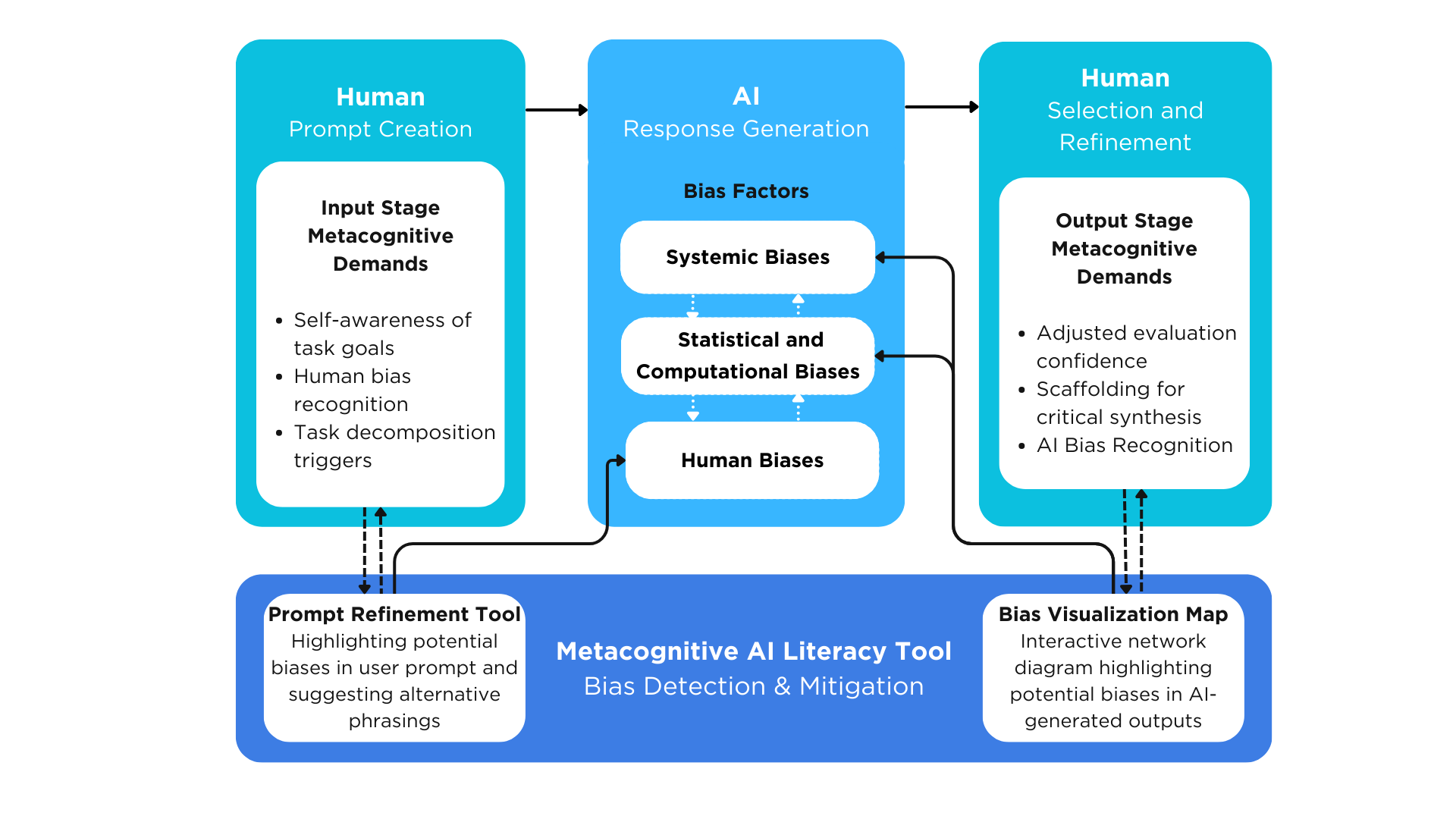}
  \caption{Bidirectional AI Literacy Intervention in Human-AI Collaborative Framework for Metacognitive Support} 
  % \Description{}
      \label{fig:collaborationWorkflow}
\end{figure}

\section{Illustrative Case: \textit{DeBiasMe}}
To demonstrate the application of the above frameworks, tthis paper presents DeBiasMe, a metacognitive AI literacy tool designed to enhance awareness of human and AI biases in both the input and output stages of human-AI collaboration (See Fig. \ref{fig:collaborationWorkflow}). 

\subsection{Identifying metacognitive support needs}
Initial mixed-methods user research with university students using surveys and interviews identified three metacognitive support needs, informing the design to meet the following learner requirements: (1) bias awareness (understanding how human biases affect prompt formulation); (2) AI understanding (comprehending AI's limitations and capabilities); and (3) critical thinking (developing self-awareness of reasoning processes during AI interactions). 

\subsection{Frictional design for metacognitive support}
\begin{figure}[h]
    \centering
    \includegraphics[width=\textwidth]{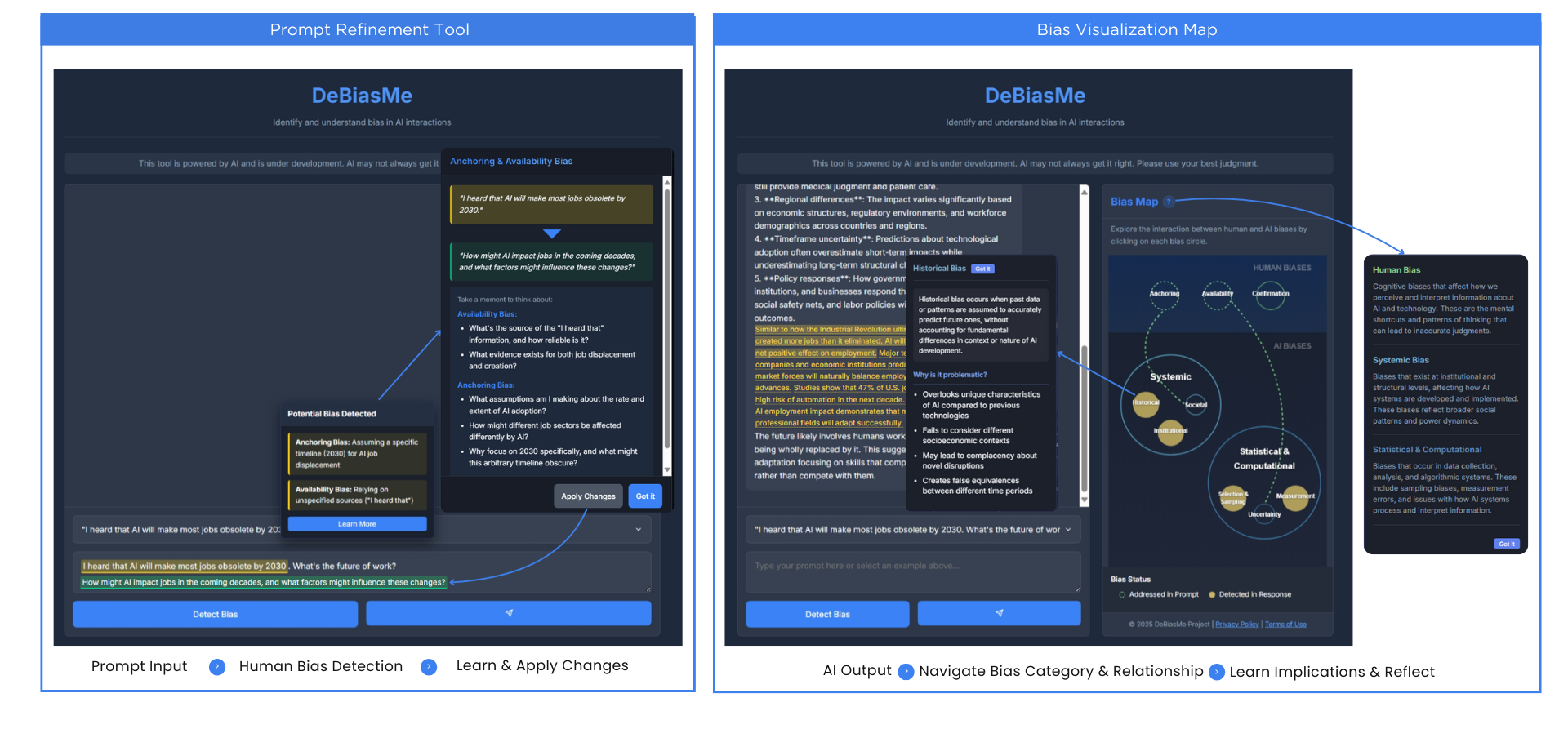} 
    \Description{}
    \caption{Interactive prototype and frictional design elements  
    (The previous UI has been fully integrated with AI by using GPT-4 on Azure)} 
    \label{fig:interactive_prototype}
\end{figure}

Based on identified metacognitive needs, design solution introduces deliberate frictions at input and output stages of the human-AI workflow through an interface that integrates two core interventions  (See Fig. \ref{fig:interactive_prototype}):

\subsubsection{Prompt Refinement Tool} Before submitting prompts to an AI system, students receive real-time feedback highlighting potential biases and suggesting alternative phrasings. This input-stage intervention helps students identify and mitigate cognitive biases before AI interaction, such as anchoring bias in how questions are framed. When activated, the tool allows users to detect bias, reflect on their implications, and apply relevant changes for prompting.

\subsubsection{Bias Visualization Map} After receiving an AI-generated response, users are presented with an interactive bias diagram highlighting potential biases in AI-generated outputs. This intervention visually maps different types of biases and their relationships by connecting information circles to the relevant text segments. By allowing users to navigate explanations of the identified bias types (human, systematic, statistical, and computational), their potential impact, and status ("addressed in prompt" and "detected in response"), the tool makes abstract bias concepts concrete and actionable. 

Through the evaluation of an earlier version of the prototype, assessing both technical effectiveness and educational effectiveness with usability heuristics, think-aloud studies, and expert reviews \cite{Nielsen1994, wang2005}, the main interventions demonstrated potential as metacognitive AI literacy tools by making implicit human and AI biases explicit and actionable, helping students develop awareness of their own thinking patterns when interacting with AI. This addresses two main components of metacognitive skills: metacognitive knowledge (enhanced awareness of biases and thought processes) and metacognitive regulation (actively monitoring and adjusting) for critical engagement with AI tools \cite{lee2025}. 

\subsubsection{Bias detection approach}
To provide real-time bias detection and visualization within the user's AI tool environment, the current bias detection approach uses AI to analyze text for potential biases against a hierarchical bias taxonomy established with main risks and vulnerabilities of human and AI system biases \cite{Schwartz2022}. Current development is further refining bias detection by incorporating both hierarchical classification of bias types and analysis of human-AI bias relationships. 

\section{Developing AI Literacy Interventions for Human Bias}
Following sections will outline ongoing development of the tool for developing AI literacy tools for human bias detection and mitigation: (1) enhancing scaffolding mechanisms for diverse users, (2) refining bias visualization techniques that effectively represent the relationship between different types of biases, and (3) developing comprehensive bias measurement frameworks for evidence-based evaluation across multiple bias dimensions. 

\subsection{Scaffolding mechanisms}
Scaffolding mechanisms of the tool can be further developed by supporting diverse engagement patterns while addressing user trust. 

\subsubsection{AI use informed by bias implications} 
Tools can help students understand the broader implications of different types of biases in academic contexts. For example, anchoring bias identification is particularly relevant when addressing the pitfalls of AI collaboration that favors automation (i.e., completing tasks by finding answers to questions) over augmentation (i.e., critically questioning the questions themselves) \cite{rastogi2020deciding}. Communicating bias implications in relation to AI capabilities and limitations can further promote ethical AI use.

\subsubsection{Developing for Specific Bias-Aware Use Cases} 
Future tools could explore two potential directions of development: (1) training instruments for bias awareness, developing metacognitive skills in controlled educational environments where biases can be safely identified with pre-defined scenarios and discussed; and (2) real-time assistants integrated into students' everyday AI use workflows, detecting and flagging biases in different task contexts. Specific scaffolding mechanisms could focus on designing different modular components to support either workflow depending on institutional, educator, and learner needs, and developing appropriate evaluation methods for each approach. 

\subsubsection{Addressing User Trust in AI for Learning Support} 
The relationship between user bias and system bias detection creates complex trust dynamics that require further examination. When users' biases align with AI system biases, users may resist or reject the tool's feedback \cite{Dzindolet2003}. When there is alignment between user and system biases, identified or flagged biases from the tool can trigger defensive reactions \cite{Bickmore2005} . While the tool is designed as an exploratory tool, determining whether certain features require compulsory engagement depends on educational needs and engagement patterns across users with varying levels of trust and agency. 

\subsection{Bias Visualization}
Different bias visualization approaches are explored to further enhance the accessibility of information for users across different levels of technical expertise. 

\subsubsection{First-hand tool for qualitative assessments} 
Our working-in-progress visualization shows different bias relationship types ("reinforced" or "reduced/persistent") based on 1) structured comparison of biased and debiased alternatives  \cite{Holstein2020} and 2) identified connections between human and AI biases (See Fig. \ref{fig:bias_visualization}). This approach can facilitate qualitative assessment of interactions between human, statistical, and systemic biases within and across different AI models, fostering responsible AI use \cite{selbst2019}. For example, students can be made aware that their confirmation bias could lead them to frame prompts in ways that exploit existing cultural biases in AI training data, resulting in outputs that further reinforce their pre-existing beliefs. In other cases, students are guided to view one AI model as one among multiple options, selecting it based on an explicit assessment of its relevance and reliability. 

\subsubsection{Progressive Visualization} Visualizations can be integrated into different steps of the user's journey when visualization elements are introduced progressively rather than all at once. This approach would allow users to first identify bias presence, then explore bias types, and finally examine implications and mitigation strategies, reducing cognitive load while constructing a mental model of the bias in human-AI interaction \cite{Chandler1991}. The role of progressive disclosure can also be further explored in terms of features enabling the adjustment of visibility of bias severity and implications for individual users. When users try to be more sensitive to certain types of biases (e.g., confirmation), allowing personalized use becomes important. However, given that this type of visualization can also facilitate uncertainty and anchoring bias, different types of visualization need to be compared \cite{Procopio2022} .

\subsubsection{Adaptive Visualization} 
Future tools need to support learners with different expertise levels, giving control in their learning pathways \cite{Bai2012}. This can be done by implementing adaptive visualization complexity, allowing users to adjust the level of detail based on their comfort with the system and the specific requirements of their tasks. A promising direction for development involves implementing task-specific visualization modes tailored to different educational contexts. For example, research-focused visualizations might emphasize potential confirmation biases in source selection and interpretation, while creative writing-focused visualizations might highlight framing biases and limitations in perspectives. This contextual adaptation would enhance the tool's relevance across various academic activities and learners.
\begin{figure}[h]
    \centering
    \includegraphics[width=\textwidth]{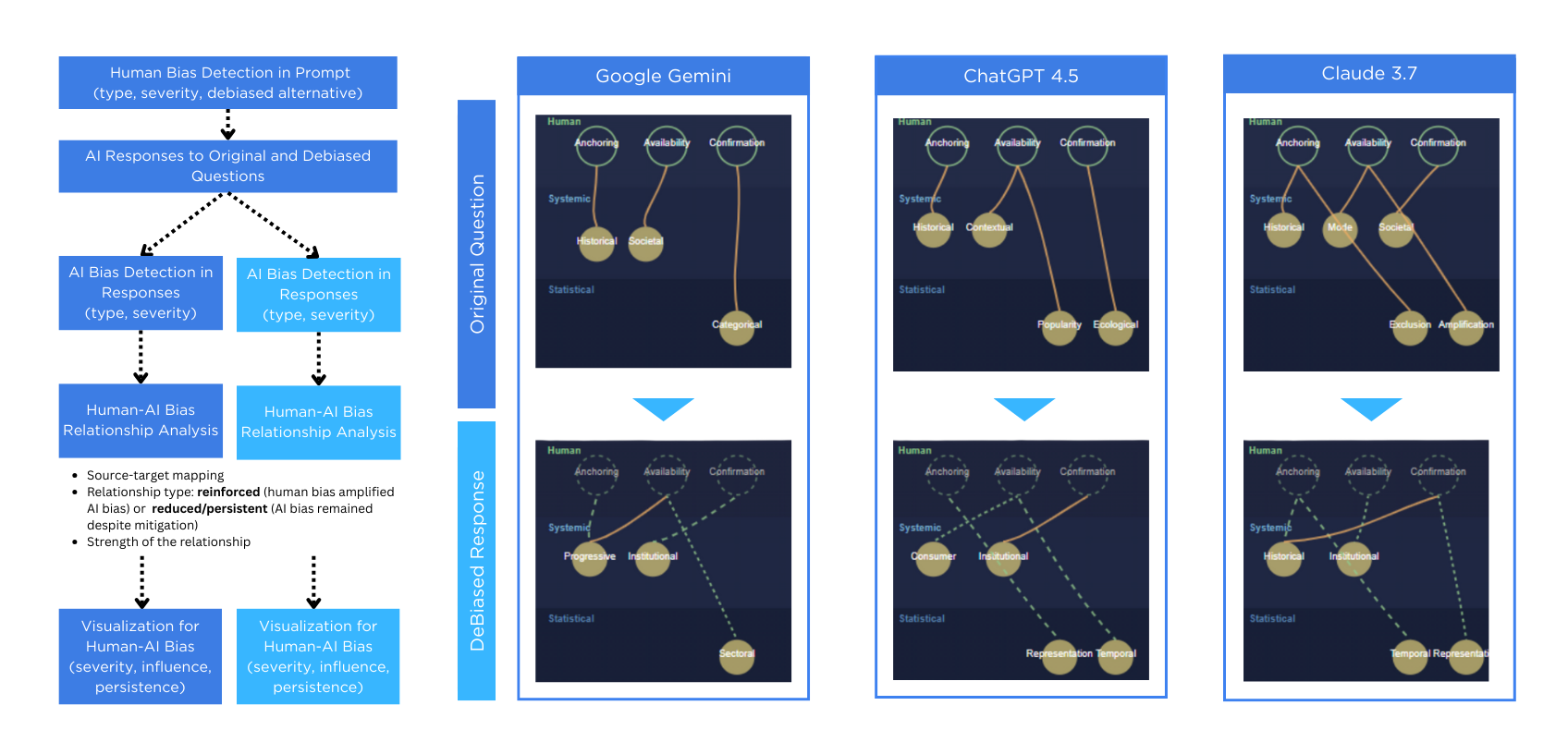} 
    \Description{}
    \caption{Developing bias visualization: A structured Comparison of Biased and Debiased Alternatives} 
    \label{fig:bias_visualization}
\end{figure}

\subsection{Bias Analysis and Mitigation Approaches}
Developing quantitative metrics for bias detection and mitigation of the tool requires further technical considerations to develop a multi-dimensional bias index framework for bias detection and mitigation. 

\subsubsection{Bias measurement (prevalence, intensity, and mitigation)} 
Mitigation of bias prevalence and intensity can be further assessed in terms of analysis accuracy and educational relevance. The success rate of bias mitigation strategies can be measured through changes in subsequent interactions after bias feedback. These metrics can be calculated both at the individual level and aggregated across user populations to identify common patterns and the effectiveness of the tool interventions from technical perspectives. This can further benefit from regular external auditing processes by experts in education and AI safety to assess both system performance and educational quality. 

\subsubsection{Methodological Integrity of Bias Analysis} 
A fundamental challenge in developing an AI literacy tool integrating bias detection functionality in AI is ensuring the methodological integrity of the detection system itself. This presents a multi-layered challenge. While AI biases reflect incompleteness of training data, biases in AI systems can also be deliberate for deception \cite{Park2024}. To address this, several safeguard methods can be considered. The tool can adopt a multi-model consensus approach that compares bias assessments from different AI systems with diverse training methodologies \cite{Binns2018}. By identifying areas of agreement and disagreement between these systems, the tool can triangulate more reliable bias assessments while highlighting areas of uncertainty. Further, implementing transparent bias analysis systems will be crucial to clearly distinguish between different levels of confidence in bias assessments. Rather than presenting all bias detections with equal certainty, the system can communicate confidence levels based on the strength of evidence and consensus across detection methods. 

\section{Discussion}
This position paper contributes to the emerging field of AI-augmented reasoning by emphasizing the critical role of metacognition in human-AI collaboration and the focus on human bias. There are several limitations that needs to be acknowledged to further develop this tool in line with the responsible innovation framework, which considers the potential societal impacts of emerging technology \cite{stilgoe2020, UKRI2023}. 

\subsubsection{Limitations} 
While this tool focused on addressing individual bias guided by self-regulation theory \cite{Azevedo2004}, biases in collaborative contexts in shared learning environments can be further examined. Further, this tool adopted an established bias taxonomy, but this can be further explored with co-designing sessions employing community engagement and creative approaches exploring bias-related themes and visualizations, data representation and data politics. Finally, while this tool focused on LLMs, future development on different types of biases, for example with image generation, is worth exploring with multimodal AI. 

\subsubsection{AI Literacy for epistemic ethics and civic imagination} 
AI literacy requires the assessment of complex affective, behavioral, cognitive, and ethical dimensions \cite{ng2024}. \textit{DeBiasMe} provides both theoretical and technical solutions by prompting students to evaluate whether AI assistance is necessary for a given task, encouraging a more reflective approach to AI use. This paradigm shifts away from treating AI as an automatic solution to problems, instead positioning it as a strategic cognitive aid that should be deployed selectively and intentionally. This approach fosters adaptive, empathetic leadership in an era of AI sovereignty, where educational institutions and individuals maintain meaningful control over their AI infrastructure, data governance, and decision-making processes \cite{Lin2025,Atenas2024}. 

\subsubsection{Educational and Social Implications} 
This approach will benefit multiple stakeholders in the educational ecosystem: At the institutional level, tools like \textit{DeBiasMe} can serve as bridges between academic integrity policies and students' everyday AI interactions, providing structured support for critical engagement while acknowledging AI's growing role in academic work \cite{Fowler2023}. At the societal level, the tool can help address the digital confidence divide by actively involving end-users in the research, design, and dissemination of AI literacy tools \cite{Bozic2023}. Enhanced transparency regarding both human and AI biases can reshape the user-data-algorithm lifecycle in AIED by making decision processes more visible, empowering user agency. This can further inspire policy recommendations with frictional design, introducing pauses in AIED products. 
 
\section{Conclusion}
AI literacy tools must evolve beyond technical competencies to incorporate metacognitive skills that address human biases in human-AI collaboration. This position paper advocates for bias-aware AI literacy tools that encourage students to critically evaluate AI-generated content rather than defaulting to AI as an automatic authority. The illustrative tool \textit{DeBiasMe} demonstrates AI literacy frameworks in promoting critical thinking in educational contexts by prioritizing: (1) deliberate friction to enhance metacognition, (2) bi-directional Human-AI interaction intervention, and (3) supporting diverse user engagement patterns to help students navigate the complex interactions between human and AI biases, offering a promising direction for future research and development to augment learning with learner-centered AIED.

\begin{acks}{Acknowledgments}
I would like to thank Manni Cheung for his valuable early input and ongoing prototype development, UCL Interaction Centre, all participants, and Matthew Haye for their insights.
\end{acks}

\bibliographystyle{ACM-Reference-Format}
\bibliography{DeBiasMe}

% %%
% %% If your work has an appendix, this is the place to put it.
% \appendix
% \pagebreak
\end{document}